\documentclass{iaus}
\usepackage{graphicx}

\title[~~Spikes \& Ripples] 
{Spikes in the SED and Ripples \\ in the Outskirts of Galaxies}

\author[Sukanya Chakrabarti]   
{Sukanya Chakrabarti$^1$}

\affiliation{$^1$777 Glades Road, Florida Atlantic University, Boca Raton, FL 33431 \\ email: {\tt schakra1@fau.edu} }

\pubyear{2011} 
\volume{284}  
\pagerange{1--12}
\jname{The Spectral Energy Distribution of Galaxies}
\editors{R.J. Tuffs \&  C.C.Popescu, eds.}

\begin{document}

\maketitle

\begin{abstract}

We describe a new method that allows us to quantitatively characterize galactic satellites from analysis of disturbances in outer
gas disks, without requiring knowledge of their optical light.  We have demonstrated the validity of this method, which we call Tidal Analysis, by applying it to local spirals with known optical companions, including M51 and NGC 1512.   These galaxies span the range from having a low mass companion ($\sim$ one-hundredth the mass of the primary galaxy) to a fairly massive companion ($\sim$ one-third the mass of the primary galaxy).   This approach has broad implications for many areas of astrophysics -- for the indirect detection of dark matter (or dark-matter dominated dwarf galaxies), and for galaxy evolution in its use as a decipher of the dynamical impact of satellites on galactic disks.   Here, we present some preliminary results on the emergent SEDs and images, calculated along the time sequence of these dynamical simulations using the 3-D self-consistent Monte Carlo radiative transfer code RADISHE.  We explore star formation prescriptions and how they affect the emergent SEDs and images.  Our goal is to identify SED colors that are primarily affected by the galaxy's interaction history, and not significantly affected by the choice of star formation prescription.  If successful, we may be able to utilize the emergent UV-IR SED of the primary galaxy to understand its recent interaction history.

\end{abstract}

\firstsection 
\section{Introduction}

The current paradigm of structure formation in the universe (\cite[White \& Rees 1978]{WhiteRees}) builds galaxies by merging smaller units, producing a universal distribution of sub-halos, with smaller haloes embedded within larger haloes on all scales.  This paradigm successfully recovers the observed large-scale distribution of galaxies (\cite[Colless et al. 2001]{Coll01}) using numerical simulations with increasing fidelity (\cite[Springel et al. 2006]{Spring06}).  However, it is not yet clear whether it applies equally well to sub-galactic scales. The excess of dark-matter dominated dwarf galaxies in dissipationless cosmological simulations of the Milky Way relative to Local Group dwarfs (\cite[Klypin et al. 1999]{Kly99}) -- dubbed the "missing satellites problem" -- can make one wonder about the applicability of the prevailing cold dark matter model on sub-galactic scales. 

We recently developed a novel method whereby one can infer the mass, and relative position (in radius and azimuth) of satellites from analysis of observed disturbances in outer gas disks, without requiring knowledge of their optical light (\cite[Chakrabarti \& Blitz 2009, henceforth CB09; Chakrabarti \& Blitz 2011, henceforth CB11; Chakrabarti, Bigiel, Chang \& Blitz 2011, henceforth CBCB; Chang \& Chakrabarti 2011, henceforth CC11]). We applied this method to M51 and inferred that its satellite has a mass  one-third that of the primary galaxy, with a pericentric approach distance of 15$~\rm kpc$.  These estimates are corroborated by observations (Smith et al. 1990) and recent simulation studies (\cite[Salo \& Laurikainen 2000]{SL00}).  Moreover, at the time when our simulations achieve the best-fit to the HI data, the azimuth of the satellite in the simulations agrees closely with its observed location.  CBCB note that the derivation of these numbers is uncertain at the factor of two level due to variations in the initial conditions of the simulated M51 galaxy, orbital inclination and orbital velocity of the satellite.  We call this method ``Tidal Analysis'' (henceforth TA).  Most recently, we have found that we can build on our earlier results to infer the scale radius of the dark matter halo in the primary galaxy itself (\cite [Chakrabarti 2011]{Chak2011}), and thereby constrain the potential of the dark halo.

Our work in this sequence of papers is motivated by the question -- can dark (or nearly dark) galactic satellites (and the dark matter density profile) be characterized from tidal gravitational effects on the outer gas disks of galaxies?  This question and our method have far-reaching implications in many areas of astrophysics. Our method is complementary to gravitational lensing in probing mass distributions without requiring knowledge of their stellar light, although it is not subject to uncertainties in the projected mass distribution (\cite[Treu et al. 2002]{Treu02}), as is lensing. It provides a means of indirect detection of dark matter dominated objects, and may be correlated with gamma ray studies (\cite[Strigari et al. 2008]{Strig08}) to hunt for dark matter dominated dwarf galaxies.  TA also offers a potential route to address the missing satellites problem.   Therefore, it may allow us to investigate whether the prevailing cold dark matter model applies equally well to sub-galactic scales. Finally, recent observations of disturbances in the outskirts of spiral galaxies (Levine, Blitz \& Heiles 2006; Thilker et al. 2007;  Bigiel et al. 2010) prompt the question whether these disturbances arise from passing galactic companions, and trigger the observed star formation in the very outskirts.

The contemporary hunt for dark matter has much in common with the hunt for planets back in the 1800s.  In 1846, Urbain Le Verrier analyzed disturbances in the orbit of Uranus.  He hypothesized that these perturbations were due to an yet unseen planet.  He was able to calculate the azimuth of the perturbing planet, now called Neptune, to within a degree.  To the best of this author's knowledge, this is the first example of the discovery of an essentially dark object from analysis of its gravitational effects on another body.  The work that we are trying to do -- to characterize dark matter dominated satellites and the dark matter density profile -- from gravitational effects on outer gas disks,  is in a similar spirit.  After the discovery of Neptune, Le Verrier became interested in understanding the perihelion precession of Mercury.  He believed it was due to a planet he called Vulcan.  Le Verrier's mistake in this case is due to his lack of investigation of the possible incidence of false positives in attributing orbital anomalies to other planetary bodies.  However, given that there was little available data at that time that he could test his theories on, his error was perhaps understandable.  

The most significant difference in the field of astronomy then and now is the explosion of data, providing theorists ample opportunity (if they are able to avail themselves of it) to test their models.  We are currently exploring the effects of intrinsic processes (such as torques arising from a non-spherical halo) to see if they can mimic tidal effects.   Preliminary work (Chakrabarti, Debattista \& Blitz, in preparation) finds that the evolution of the halo shape (and thereby the strength of the Fourier amplitudes in the outskirts of the gas disk as determined by intrinsic processes) is significantly affected by gas cooling and angular momentum transport from the gas to the halo.  The shapes of halos are considerably rounder close to present day, when the stellar to dark matter masses are comparable to local spirals ($M_{\star}/M_{\rm DM} \sim 0.03$) (\cite[Leroy et al. 2008]{Leroy08}), relative to $z \sim 2$.  Examination of the Fourier amplitudes of these simulations close to present day indicates that the Fourier amplitudes in the outskirts are $< 10 \%$, which suggests that non-spherical halos would not be the primary contributor to the observed strength of the Fourier amplitudes in local spirals.  This is certainly valid for M51, which is known to be a tidally interacting system (CBCB; Salo \& Laurikainen 2000).  Local spirals that display small ($< 10 \%$ in the Fourier amplitudes, as defined in CB09, relative to the axisymmetric mode) perturbations in the outskirts, and have no visible tidally interacting companion may require a more careful treatment of the effect of halo shapes. 

We are currently in the process of building on our earlier results to explore whether the emitted SED can also be used to independently characterize galactic satellites.  An extensive multi-wavelength (from the radio to the x-ray) database exists for local spirals, e.g., built around the SINGS (Spitzer Infrared Nearby Galaxy Survey) project (\cite[Kennicutt et al. 2003]{Kenn03}).  In addition to generating simulations that match the observed HI data, calculating the SEDs of these simulated galaxies and comparing with multi-wavelength data gives us another observational handle, in addition to the HI maps, to attest to the use of these simulations as prototypes in modeling local spirals.  Examining relations between commonly used star formation tracers (such as the $24~\mu \rm m$ flux) and the emitted luminosity will also be useful in interpreting observations of high redshift galaxies.

It is now well known that major mergers have higher FIR/UV flux than secularly evolving galaxies (\cite[Chakrabarti et al. 2008]{Chak08}).  What we would like to understand here is if SED colors (such as the FIR/UV flux) can be a tracer of tidal interactions for minor mergers also (and down to how low a satellite mass).  For this tracer to be useful, the colors would need to be robust to the choice of star formation (SF) prescription, dust model (or ISM decomposition).  This work builds on our earlier work on calculating the emergent SEDs from SPH simulations (Chakrabarti et al. 2007; 2008; Chakrabarti \& Whitney 2009).  The choice of ISM decomposition (Jonsson et al. 2010; Narayanan et al. 2010) will affect the emergent SED, and we are exploring a range of prescriptions to understand the primary dependences.  We have also recently implemented a model of PAH emission following Wood et al. (2008).  Particularly massive satellites may induce some star formation in the outskirts of galaxies.  Therefore, we are exploring a range of SF prescriptions, including works that have treated the effects of shock heated star formation (Barnes 2004), pressure (Blitz \& Rosolowsky 2006), and molecular gas fraction (Krumholz et al. 2009), to understand the effect of varying star formation prescriptions on the SED.  Our goal here is to identify SED colors that are primarily affected by the galaxy's interaction history, and not significantly affected by the choice of SF prescription.  
This study may provide a way to diagnose the strength of interactions in high-redshift galaxies (many of which are thought to be produced by minor mergers; Genzel et al. 2010) that are being observed by $\it{Spitzer}$ and $\it{Herschel}$, but cannot be mapped adequately in HI.

\section{Proof of Principle of Tidal Analysis:  M51}

\begin{figure}
\begin{center}
\includegraphics[scale=0.45]{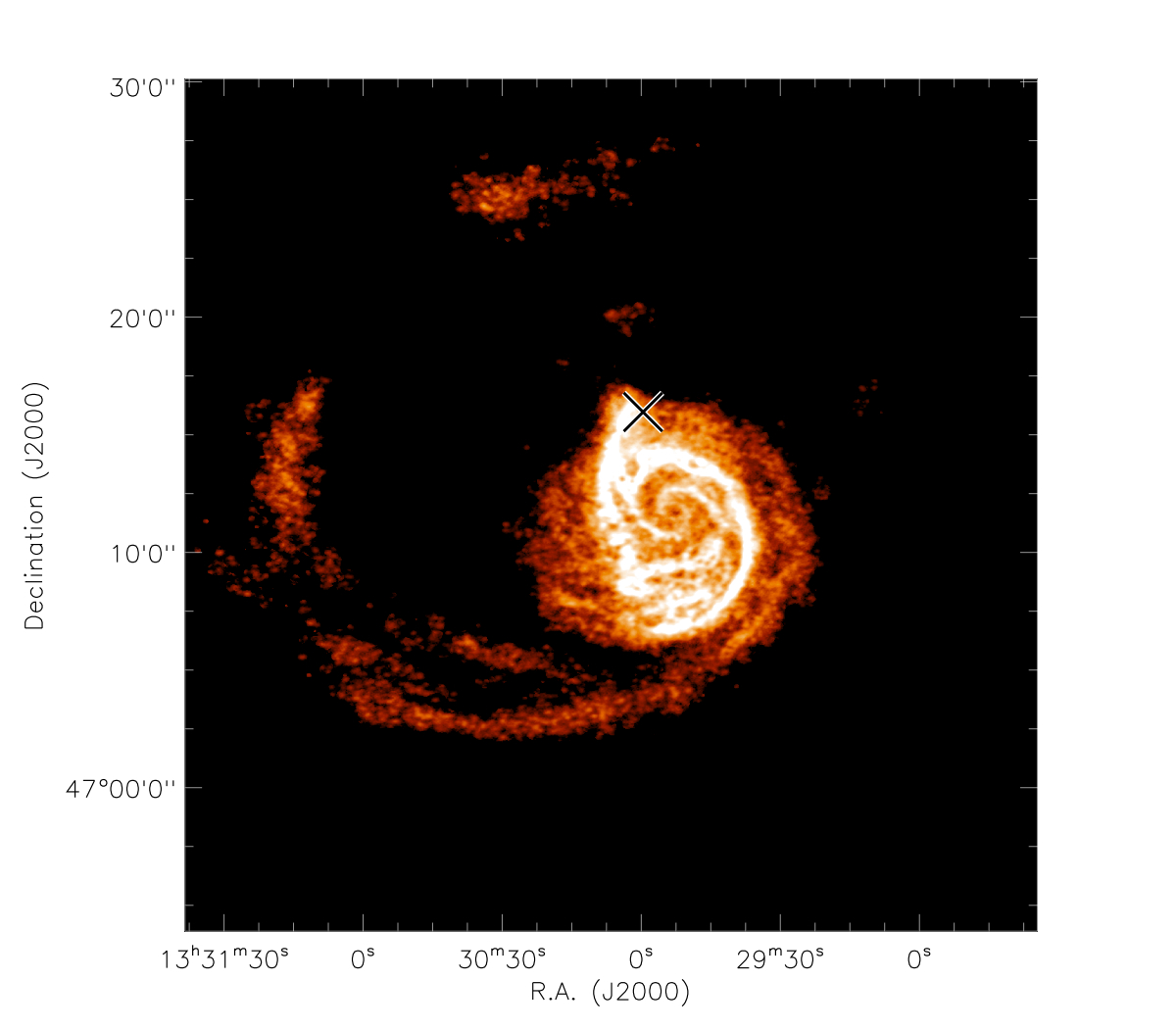}
\caption{THINGS VLA image of M51 showing the HI distribution.  Note that the companion of M51 sits at the short arm, as marked by the cross.  Adapted from Chakrabarti et al. (2011) \label{f:m51data}}
\end{center}
\end{figure}

\begin{figure}
\begin{center}
\includegraphics[scale=0.3]{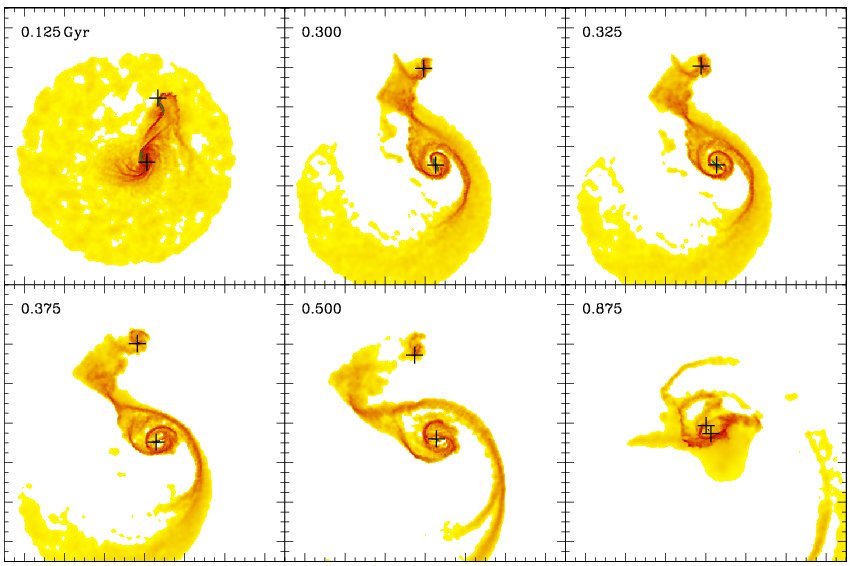}
\caption{Gas density images of best-fit simulation of M51, with crosses marking centers of both galaxies.  The best-fit time to the Fourier amplitudes occurs at $t \sim 0.3~\rm Gyr$.  The box extends from -80 kpc to 80 kpc.  Adapted from Chakrabarti et al. (2011).  \label{f:m51sim}}
\end{center}
\end{figure}

The simulations we discuss here have the same setup as in CBCB.  We carry out SPH simulations using the GADGET code (\cite[Springel (2005)]{Spring05}), of M51 interacting with its companion.  CBCB carried out a simulation parameter survey and compared the resultant Fourier amplitudes of the low order modes of the gas surface density with the observed HI data of M51.  They found that placing simulations on a variance vs variance plot (where the variance is  with respect to the low order modes of the simulations and the data) made the best-fit simulations visually apparent.  Specifically, CBCB found that the best-fit to the HI data occurred for a 1:3 mass ratio satellite with a pericentric distance of $15~\rm kpc$, parameters that are corroborated observationally and are in agreement with other simulation studies (Smith et al. 1990; Dobbs et al. 2010; Salo \& Laurikainen 2000).  CBCB also found that the azimuthal location of the companion of the best-fit simulation agrees very closely with the observed location of M51's companion, at the time when the Fourier amplitudes most closely match the data.  Figure \ref{f:m51data} shows the HI map of M51, with the companion's location marked by the cross.  Figure \ref{f:m51sim} shows the gas density images of the best-fit simulation.  The best-fit time to the Fourier amplitudes occurs at $t \sim 0.3~\rm Gyr$, when the satellite's location lies close to the tip of the short arm, as it does in the real galaxy.  CBCB also found that the method was successful in characterizing NGC 1512's satellite ($\sim 1:100$ mass ratio), which is a much lower mass satellite than M51's relatively massive satellite.  Thus, CBCB concluded that analysis of observed disturbances in the extended HI disks of galaxies can be used to infer the mass and current distance (in radius and azimuth) of galactic satellites.  Earlier work in this series of papers presented the basic reasoning as to why the mass-pericentric approach degeneracy in the tidal force can be broken when the time integrated response of the primary galaxy is considered (CB09), and described the method to find the azimuth of galactic satellites from the phase of the modes (CB11).  The reason why we focus our analysis on the gaseous disk is that disturbances in the gas disk dissipate on the order of a dynamical time -- leaving a clean slate, and therefore allow an easier interpretation of satellite interactions than the stellar disk, where past interactions are still visible after a dynamical time.

\subsection{Exploring Star Formation Prescriptions}

We begin by contrasting the effects of two different star formation prescriptions, namely the Kennicutt-Schmidt star formation prescription and the Barnes shock heated star formation prescription.  The former depends only on the density (the implementation in GADGET is described in Springel et al. 2005), while the latter depends on the density and the time rate of change of entropy. Barnes (2004) proposes the following form for the probability of gas particle $i$ undergoing star formation in time $\Delta t$:
\begin{equation}
p_{i}=C_{\star}  \rho^{n-1}  \rm MAX(\dot{u} ,0)^{m} \Delta t 
\end{equation}
where $u$ is the entropy and the MAX function, which returns the larger of its two arguments, is used to handle cases where $\dot{u}$ is less than zero.  Shocks are identified as regions with $\dot{u}$ significantly larger than zero.  Thus, setting $m=0$ and $n > 1$ yields a density dependent prescription (with $n=2.5$) yielding Kennicutt-Schmidt, and setting $m>0$ yields shock heated star formation.  Here,
we set $m=1$ and $n=2.5$.

\section{Emergent SEDs \& Images}

We use the self-consistent 3-D Monte Carlo radiative transfer code RADISHE to calculate the emergent SEDs and images through the time outputs of the SPH simulations, as in prior work (Chakrabarti et al. 2007; 2008; Chakrabarti \& Whitney 2009).  The radiative transfer methodology is described in Chakrabarti \& Whitney (2009; henceforth CW09).  Since dust envelopes can in principle be optically thick to their own reprocessed emission, it is necessary to calculate the temperature in a global way, i.e., the optically thin approximation is not valid in general.  This is particularly true for dusty galaxies like ULIRGs.  We described the implementation of the dust temperature calculation in CW09.  CW09 found that the Lucy (1999) temperature calculation algorithm is very efficient for large 3-D grids, and described rules of thumb for its implementation.  PAH emission is modeled following Wood et al. (2008), by sampling precomputed PAH emissivity files for a wide range of values of the mean intensity of the radiation field.   

\begin{figure}
\begin{center}
\includegraphics[scale=0.35]{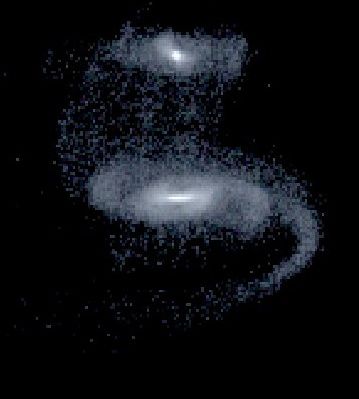}
\includegraphics[scale=0.35]{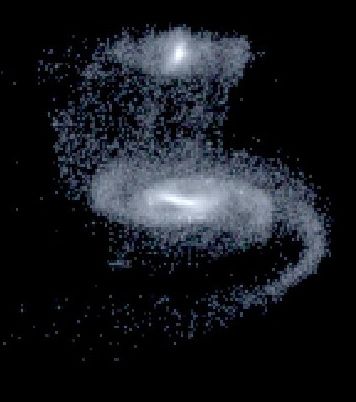}
\caption{Simulated near-IR images of M51, using the (a)Kennicutt-Schmidt prescription, and (b) the shock heated star formation prescription.   \label{f:m51images}}
\end{center}
\end{figure}

\begin{figure}
\begin{center}
\includegraphics[scale=0.5]{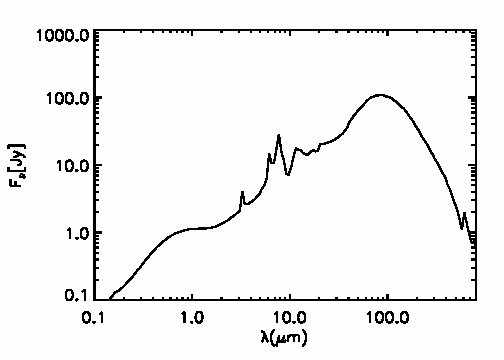}
\caption{Simulated SED of M51 showing clear PAH features.  \label{f:m51sed}}
\end{center}
\end{figure}

Figure \ref{f:m51images} (a,b) displays simulated images of M51 close to the best-fit time in the J, H, K bands.  Using the Kennicutt-Schmidt prescription (a),
yields lower near-IR flux from the tidal bridge than when the shock heated star formation prescription is employed (b).  This is to be expected as the Barnes (2004) 
prescription takes both the density (which is low in tidal tails overall) and shocks (which do occur in tidal tails) into account, while the K-S prescription only accounts
for the density.  Figure \ref{f:m51sed} shows the emergent SED (at the time that best fits the Fourier amplitudes of the HI data).  Copious PAH emission features are present in the mid-IR, which we find commonly emerge in tidally
interacting simulations.  We are in the process of analyzing SEDs of these simulations to search for trends that may allow us to infer the galaxy's interaction history
from the SED.

\section{Summary \& Future Work}

In summary, we find that the dynamical echoes of the last generation of galactic satellites on the fragile extended HI disks of spirals can be analyzed
to learn a number of things about dwarf galaxies (their mass and location), without requiring any knowledge of their optical light.  This method, which we call Tidal Analysis, 
gives us a means of mapping the dark matter distribution.  In this way it is analogous to gravitational lensing.  Most recently, we have used the radiative transfer code RADISHE (Chakrabarti et al. 2008) to calculate emergent SEDs and images through the time outputs of these simulations of locally interacting spirals.  The goal is to utilize the emergent UV-IR SED of the primary galaxy to understand its interaction history, in addition to the HI map.  Future work includes studying the effects of a range of dust models, and more realistic star formation prescriptions  to understand their effects on the emergent SED.  The dynamical models and the Tidal Analysis method will be tested by applying it to the large THINGS sample to determine its statistical viability.  Cross comparison of the SEDs and images against the data from the SINGS survey will allow us to test the radiative transfer models and implicit assumptions therein.

\end{document}